\documentclass{article}
\usepackage{spconf,amsmath,graphicx}
\usepackage{listings}
\usepackage{color}
\usepackage{multirow}
\usepackage{booktabs}
\usepackage{float}
\usepackage{enumitem}
\usepackage{tabularx}
\usepackage{multirow}
\usepackage{makecell}
\usepackage{bm}

\title{Dynamic Chunk Convolution for Unified Streaming and Non-Streaming Conformer ASR}
\name{Xilai Li, Goeric Huybrechts, Srikanth Ronanki, Jeff Farris, Sravan Bodapati}
\address{AWS AI Labs\\
\small{\{lixilai, huybrech, ronanks, jjfarris, sravanb\}@amazon.com}}
\begin{document}
\ninept
\maketitle
\begin{abstract}

Recently, there has been an increasing interest in unifying streaming and non-streaming speech recognition models to reduce development, training and deployment cost. The best-known approaches rely on either window-based or dynamic chunk-based attention strategy and causal convolutions to minimize the degradation due to streaming. However, the performance gap still remains relatively large between non-streaming and a full-contextual model trained independently. To address this, we propose a dynamic chunk-based convolution replacing the causal convolution in a hybrid Connectionist Temporal Classification (CTC)-Attention Conformer architecture. Additionally, we demonstrate further improvements through initialization of weights from a full-contextual model and parallelization of the convolution and self-attention modules. We evaluate our models on the open-source Voxpopuli, LibriSpeech and in-house conversational datasets. Overall, our proposed model reduces the degradation of the streaming mode over the non-streaming full-contextual model from 41.7\% and 45.7\% to 16.7\% and 26.2\% on the LibriSpeech \textit{test-clean} and \textit{test-other} datasets respectively, while improving by a relative 15.5\% WER over the previous state-of-the-art unified model. 
\end{abstract}

\section{Introduction}
\label{sec:intro}

End-to-end (E2E) automatic speech recognition (ASR) models such as attention-based encoder-decoder (AED) \cite{chan2015listen, chorowski2015attention}, CTC \cite{graves2006connectionist, amodei2016deep} and Transducer \cite{graves2012sequence, graves2013speech, dalmia2021transformer} have gained a lot of attention over the past decade due to their simplicity in the integration of the pronunciation, language and acoustic models into a single neural network.
While state-of-the-art E2E models work remarkably well in a non-streaming fashion, they suffer from degradation when operating in a streaming manner as the requirement of transcribing text in real time poses an extra challenge. Numerous works try to bridge the gap with non-streaming ASR by training a model specific for a streaming ASR task \cite{sainath2019two, chiu2017monotonic}, focusing on mitigating the trade-off between latency and accuracy. While some slight improvements have been observed, the gap with models that take the full acoustic sequence as input (a.k.a. full-contextual models) remains non-negligible \cite{zhang2020transformer, huang2020conv}.

In recent years, efforts have been made to unify  streaming and non-streaming into a single model \cite{tripathi2020transformer, zhang2020unified, yu2020dual, yao2021wenet, moritz2021dual, liu2022learning}, which helps reduce development, training and deployment cost. A commonly explored solution is to expose the unified model to various contexts at training time thereby making the model less susceptible to accuracy degradation at inference time under different latency conditions. In \cite{yao2021wenet}, a dynamic chunk training technique is adopted where the input is split into several fixed sized chunks and the audio frames within each chunk attend on themselves and frames from all the previous chunks.
They vary the chunk size dynamically from 1 to the maximum utterance length in the batch, so the trained model learns to predict with arbitrary chunk size. \cite{zhang2020unified} and \cite{kim2021multi} present a quite similar dynamic chunk-based attention strategy, but other methods exists too. \cite{narayanan2021cascaded} for instance, first processes input features with a streaming encoder before passing these to a non-streaming encoder. A single decoder then learns to decode either using the output of the streaming or the non-streaming encoder. \cite{yu2020dual} introduced dual-mode ASR with shared weights for both streaming and full-context speech recognition to further optimize the performance of streaming ASR. Similarly, the Dual Causal/Non-causal (DCN) self-attention network proposed in \cite{moritz2021dual} processes two sequences of causal and non-causal frames in parallel and prevents the overall context to grow beyond the look-ahead of a single layer. The authors in \cite{liu2022learning} employ self-supervised pre-training with wav2vec 2.0 \cite{baevski2020wav2vec} and fine-tune the model through dual-mode training. 

In general, streaming and non-streaming ASR systems are trained independently for optimized performance. These recent works on unified ASR are a great step towards a single, easy-to-use solution regardless of the inference mode. However, we identify two main shortcomings in the literature: First, the performance gap between streaming and non-streaming of an unified model still remains significant, especially when a Conformer encoder is used \cite{zhang2020unified, moritz2021dual}. Second, the gap between non-streaming and a full-contextual model enlarges with the increase in the amount of training data \cite{zhang2020unified, yao2021wenet}. 

In this work, we propose a \textit{dynamic chunk convolution} (DCConv), a non-causal convolution with an improved training strategy for unified non-streaming and streaming Conformers \cite{gulati2020conformer}. This builds further upon the dynamic chunk training (DCT) from \cite{yao2021wenet}, in which the core idea is to divide the input into chunks with a chunk size that gets dynamically generated at training time. The difference lies in our novel convolution which better mimics the inference conditions at training time while keeping a rich acoustic representation and therefore results in superior performance. Besides the proposed DCConv, we extend the original DCT with two other key contributions: a) we demonstrate a better performance in both streaming and non-streaming when the model is fine-tuned from a baseline full-contextual model; b) we further optimize the streaming performance by parallelizing the convolution and self-attention modules within each Conformer block. An extensive ablation study is performed varying chunk size, overlapping chunk ratio and left context size. Empirical evaluations measured on Voxpopuli showcase the efficacy of our proposed approach in terms of the accuracy vs latency trade-off. Overall, the proposed model achieves an average relative improvement of 15.5\% WER over the previous state-of-the-art \cite{yao2021wenet} and obtained an absolute WER of 2.0\% and 2.4\% on the LibriSpeech \textit{test-clean} dataset in the non-streaming and streaming mode, respectively. 

\begin{figure}[t]
    \centering
    \includegraphics[width=0.75\linewidth]{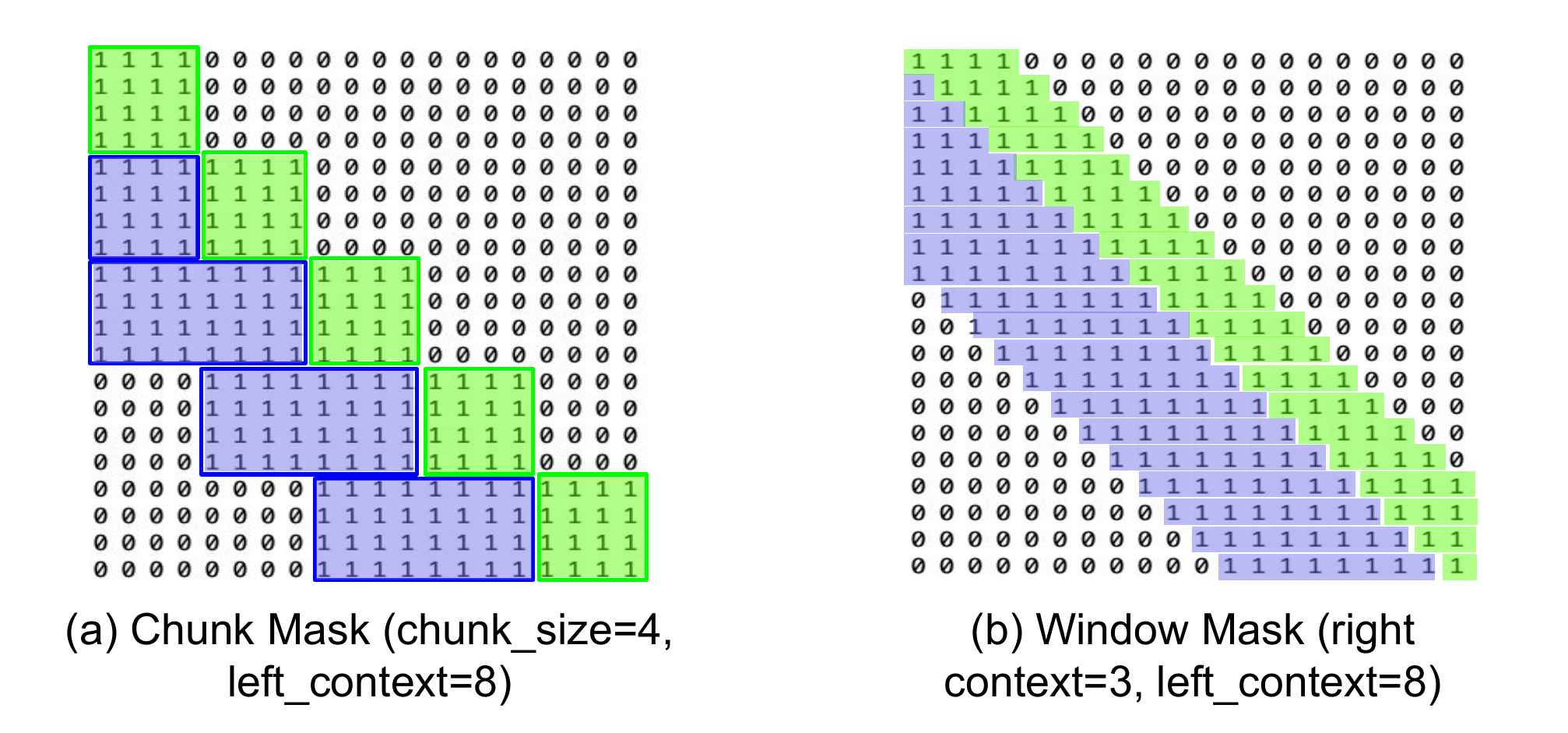}
    \vspace{-3mm}
    \caption{\small (a) An example of a chunk mask of size 4, left context size 8 and sequence length 20. (b) An example of a window mask of right context size 3, left context size 8 and sequence length 20.}
    \label{fig:attention_mask}
    \vspace{-5mm}
\end{figure}

\section{Approach}
\label{sec:approach}

\vspace{-3mm}
\subsection{Model architecture}
\vspace{-1mm}

We consider a joint CTC-Attention framework \cite{kim2017joint, dingliwal2023personalization} for training our unified models. It consists of three components: a \textit{Shared Encoder}, a \textit{CTC Decoder} and an \textit{Attention Decoder}. For our experiments, we only use the \textit{CTC Decoder} at inference time and therefore we configure the \textit{Attention Decoder} with a shallow transformer \cite{vaswani2017attention}. For the \textit{Shared Encoder}, we consider two variants: the Conformer architecture \cite{gulati2020conformer} and a parallel Conformer \cite{peng2022branchformer}. The Conformer architecture consists of a regular Conformer encoder block \cite{gulati2020conformer} in which the convolution module appears after the multi-head self-attention (MSA). We demonstrate that using a parallel Conformer (P-Conf) encoder block \cite{peng2022branchformer} instead, which has the convolution and MSA existing next to each other, is beneficial for the unified streaming scenario. The advantage of the P-Conf is to capture both global and local context explicitly in the MSA and convolution branch respectively. As opposed to the recently proposed Branchformer \cite{peng2022branchformer}, which also uses the parallel structure, we leverage the block for the streaming application too. The P-Conf reduces the overall receptive field due to its parallel nature while maintaining the same model capacity, resulting in more robust streaming performance.

\vspace{-3mm}
\subsection{Dynamic Chunk Training for Self-Attention}
\label{ssec:dct}

For unified models to perform well, they must be exposed to both limited and full context during their training. To accomplish this, \cite{yao2021wenet} propose dynamic chunk training (DCT) for self-attention layers. 
The DCT idea involves varying the chunk size dynamically from 1 to the max utterance length for different batches in training. This is achieved by applying a dynamic chunk mask to the attention score matrix for each self-attention layer, which is illustrated in Eq. 1:

\vspace{-5mm}
\begin{align}
    \text{Attn}(\bm{Q},\bm{K},\bm{V}) &= \text{Softmax}(\textbf{Mask}(\bm{Q}\bm{K}^T)/\sqrt{d})\bm{V}
\end{align}
\vspace{-5mm}

\noindent where $Q$, $K$, $V$ and $d$ denote the queries, keys, values and embedding dimension  respectively. Unlike the window mask (Fig.~\ref{fig:attention_mask}b), the chunk mask (Fig.~\ref{fig:attention_mask}a) strictly enforces the look-ahead size by setting the chunk size, while the receptive field with window masking grows linearly with the stacking of more layers. In this work, we randomly sample the chunk size between 8 (=320ms) and 32 (=1280ms) frames and the left context size between 0 and all left chunks, so that the model becomes robust to numerous sizes at inference time.

\vspace{-3mm}
\subsection{Dynamic Chunk Convolution}
\vspace{-1mm}

\begin{figure}[t]
    \centering
    \includegraphics[width=0.75\linewidth]{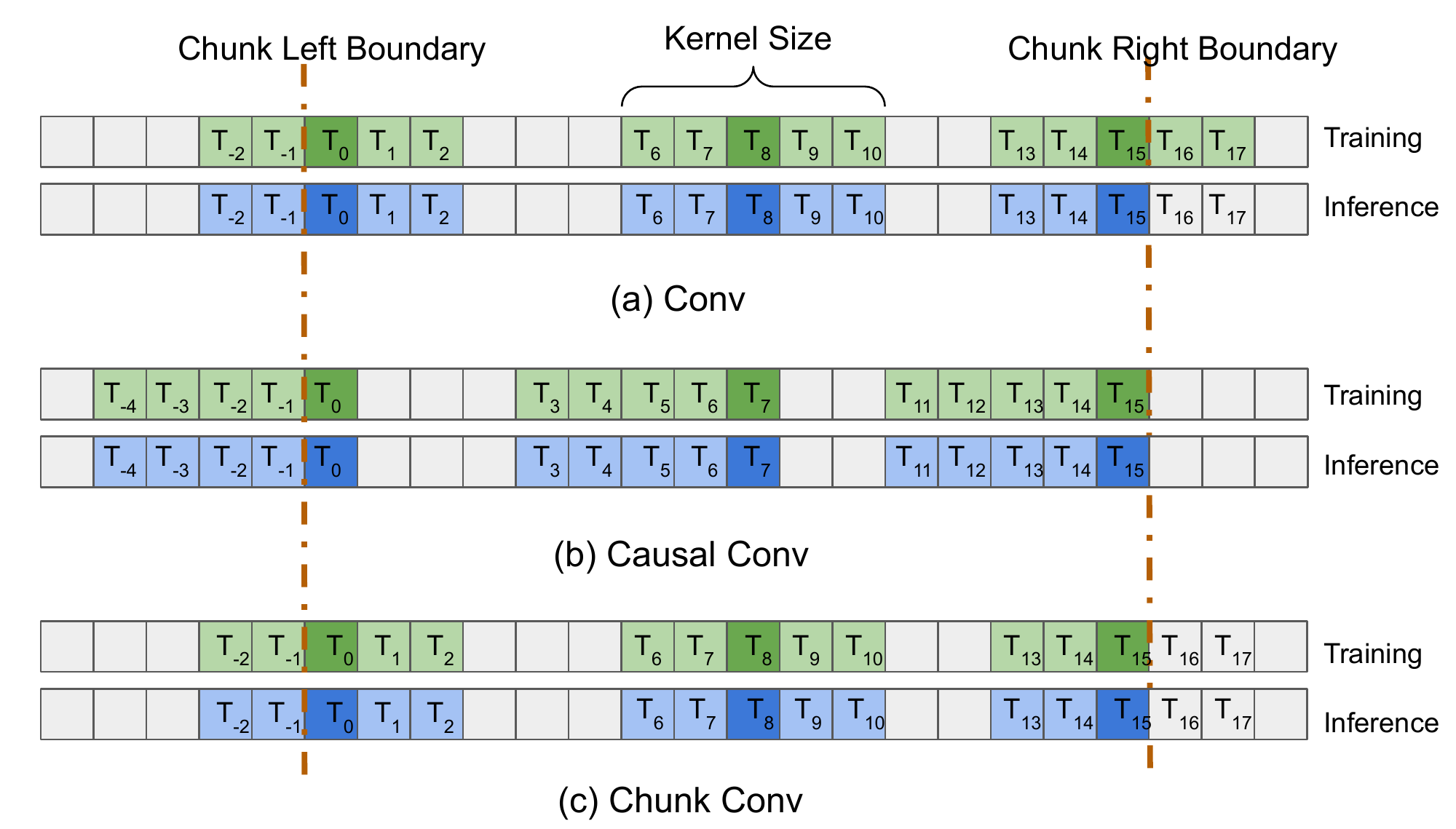}
    \vspace{-2mm}
    \caption{\small Different types of convolutions: (a) regular convolution, (b) causal convolution, and (c) chunk convolution, with example kernel size 5 and chunk size 16.}
    \label{fig:DCConv}
    \vspace{-4mm}
\end{figure}

The convolution operator is a key component of Conformer ASR models \cite{gulati2020conformer}. However, the conventional convolution results in significant accuracy degradation due to the mode mismatch between training and inference, as shown in Fig.~\ref{fig:DCConv}a. Indeed, the chunk's rightmost frames can see context from the next chunk on their right during training. Whereas at inference, this right chunk context is not available, causing a discrepancy. This inter-chunk correlation is even more magnified when stacking more Conformer blocks.

One solution that is adopted in \cite{yao2021wenet} consists of using a causal convolution, as shown in Fig.~\ref{fig:DCConv}b. The left-shifted convolution kernel restricts its span from having access to any frames beyond the chunk's right boundary. This still leads to performance degradation though, as the lack of within-chunk future context for the frame being processed results in a poorer acoustic representation.

In this work, we therefore propose a novel non-causal \textit{dynamic chunk convolution} (Fig.~\ref{fig:DCConv}c). As opposed to the conventional convolution (Fig.~\ref{fig:DCConv}a), the chunk convolution operator has no access to any future context beyond its right boundary. This trick allows training to more closely match the streaming inference setting where no future context beyond the right chunk boundary is available either. As opposed to causal convolution (Fig.~\ref{fig:DCConv}b), the DCConv chunk has access to a limited within-chunk future context of the current frame. This extra within-chunk future context results in more accurate acoustic modeling and therefore better overall accuracy. The authors of \cite{shi2022streaming} introduce a similar non-causal convolution. Unlike their convolution though, ours caches the output of the preceding chunk(s) and pads it to the current chunk, resulting in a superior representation. The non-causal convolution of \cite{shi2022streaming} is also used in the Emformer \cite{shi2021emformer} architecture for streaming use-cases with a fixed chunk size, while ours is implemented in a Conformer architecture with DCT and is applicable to both streaming and non-streaming ASR. These distinctions enable our model to be utilized in a wider range of settings.

\vspace{-5mm}
\begin{align}
    \bm{X}^{i}_{C} &= \bm{X}_{[iC-L:(i+1)C]} = [\bm{X}_{[iC-L:iC]}, \bm{X}_{[iC:(i+1)C]}] \\
    \bm{X}^{i'}_{C} &= \text{Conv}(\bm{X}^{i}_{C}) \rightarrow \bm{X}^{'} = \text{Concat}(\bm{X}^{i'}_{C [L:]})
\end{align}
\vspace{-5mm}

As shown in Eq. 2-3, we implement the DCConv by splitting the input sequence $\bm{X}$ into chunks $\bm{X}^{i}_{C}$ where $i$ denotes the index of the chunk within the sequence and $C$ the chunk size. Each chunk has a left context size $L = (kernel\_size - 1) / 2$. After the convolution is applied on every chunk, we concatenate $\bm{X}^{i'}_{C}$ from which we have removed the first $L$ output frames that correspond to the input left context. This DCConv operator does not slow down the training since all the chunks are independent from each other. Furthermore, we ensure to synchronize the size of both the chunk mask for the self-attention layers and for the DCConv such that the overall look-ahead size of the encoder is strictly set to the specified common size.

\begin{table*}[t]
    \centering
    \caption{\small Small- and Large-Scale experiments with different architectures and training strategies. All streaming evaluations are done with a 640ms chunk size, 50\% overlapping, 1280ms left context and averaged encoder latency of roughly 480ms.
    }
    \vspace{0mm}
    \resizebox{1.0\textwidth}{!}{
    \begin{tabular}{llcccc|cccc}
    \toprule
    &  & \multicolumn{4}{c}{\textbf{Relative WER}} & \multicolumn{4}{c}{\textbf{Absolute WER}} \\
    \cmidrule{3-10}
     &  & \multicolumn{2}{c}{Conversational} & \multicolumn{2}{c}{Multi-accent} & \multicolumn{2}{c}{WSJ} & \multicolumn{2}{c}{Voxpopuli} \\
    Model & Training Strategy & Non-streaming & Streaming & Non-streaming & Streaming & Non-streaming & Streaming & Non-streaming & Streaming \\\midrule
     \multirow{2}{*}{Transformer} & (A) Full-context  & - & - & - & - & 7.1 & 9.7 & 13.9 & 18.2 \\
     & (B) DCT & -11.0\% & 10.2\% & -4.2\% & 6.9\% &  7.8 & 8.9 & 14.7 & 17.1 \\\cmidrule{1-10}
     
     \multirow{5}{*}{Conformer} & (C) Full-context  & \textbf{16.0\%} & 11.6\% & \textbf{10.4\%} & 10.9\% & \textbf{6.1} & 8.5 & 12.5 & 16.1 \\
     & (D) DCT & 14.0\% & 14.3\% & 8.9\% & 12.6\% & \textbf{6.1} & 7.6 & \textbf{12.4} & 15.6 \\
     & (E) DCT w/ Causal Conv & 6.0\% & 23.1\% & 0.5\% & 8.1\% & 7 & 9.2 & 15 & 18.9 \\
     & (F) DCT w/ DCConv & 9.0\% & 27.2\% & 2.6\% & 18.6\% & 6.5 & 7.3 & 13.1 & 14.2 \\
     & (G) \hspace{0.5em} + Fine-tune & \textbf{16.0\%} & 29.3\% & 9.9\% & 21.9\% & 6.4 & 7.2 & \textbf{12.4} & 14.0 \\\cmidrule{1-10}

     \multirow{3}{*}{P-Conformer} & (H) Full-context & \textbf{16.0\%} & 14.3\% & \textbf{10.4\%} & 12.6\% & 6.3 & 8.3 & \textbf{12.4} & 16 \\ 
     & (I) DCT w/ DCConv & 8.0\% & 27.9\% & 2.1\% & 19.4\% & 6.5 & 7.3 & 13.0 & 14.0 \\
     & (J) \hspace{0.5em} + Fine-tune & \textbf{16.0\%} & \textbf{30.6\%} & 9.9\% & \textbf{22.3\%} & 6.2 & \textbf{6.9} & 12.6  & \textbf{13.8} \\\midrule\midrule
     \multirow{2}{*}{Conformer-Large} & (K) Full-context  & - & - & \textbf{-} & - & \textbf{4.5} & 6.2 & 9.2 & 12 \\
     & (L) DCT w/ DCConv + Fine-tune & \textbf{1.9\%} & \textbf{22.9\%} & \textbf{0.0\%} & \textbf{11.0\%} & 4.6 & \textbf{5.6} & \textbf{9.1} & \textbf{10.5} \\
     \bottomrule
    \end{tabular} }
    \\ [1ex]
    \label{table:5k} 
    \vspace{-4mm}
\end{table*}

\vspace{-3mm}
\subsection{Fine-tuning Baseline Full-Contextual Model}
\vspace{-1mm}

Lastly, we showcase that fine-tuning from a baseline full-contextual model results in better overall performances. Instead of training a unified model from scratch, we initialize the weights from a pre-trained full-contextual model. This approach allows to leverage the non-streaming performance of the full-contextual model. Simultaneously, the model will perform better in the streaming mode too as it can transfer common speech recognition knowledge gained from the non-streaming pre-training.

\section{Experimental Settings}
\label{sec:exp}

\vspace{-3mm}
\subsection{Datasets}

\noindent \textbf{Training data:} We consider 3 different speech corpora varying in size for training our models: A \textit{large-scale} 50k+ hour English corpus and a \textit{small-scale} 5k hour subset, sampled from in-house paired audio and text data. Both corpora include audio files with a good mix of accents, speakers, sampling rates and background noise. The third dataset is the open-source \textit{LibriSpeech} \cite{panayotov2015librispeech} corpus, for which we combine \textit{train-clean-100}, \textit{train-clean-360} and \textit{train-other-500} to have 960 hours of training data. These 3 data regimes are representative of a wide range of end-to-end ASR systems for various speech applications.

\vspace{2mm}
\noindent \textbf{Benchmarking:} For the LibriSpeech experiments in section~\ref{sec:librispeech}, we evaluate our models on \textit{test-clean} and \textit{test-other}. For the small- and large-scale experiments in sections~\ref{sec:5k} and \ref{sec:50k} respectively, we use the following test sets: (1) \textit{Conversational}: A 10+ hour in-house dataset with utterances resembling user inputs to goal oriented conversational dialog systems. The average utterance length is roughly 10 words; (2) \textit{Multi-accent}: A 100+ hour in-house long-form audio dataset, composed of 12 different accents spoken across the US. The average utterance length is roughly 16 words after segmentation; (3) \textit{Wall Street Journal (WSJ)}: We use WSJ's eval\_test92 \cite{garofolo1993csr}, prepared using Kaldi's \cite{povey2011kaldi} WSJ recipe. The dataset is 0.7h long. The average utterance length is 16 words; (4) \textit{Voxpopuli} \cite{wang2021voxpopuli}: We use the English test partition, which is 4.9h long. The average utterance length is 24 words. We report absolute word error rate (WER) on WSJ and Voxpopuli and relative WER (WERR) on in-house datasets.

\vspace{-3mm}
\subsection{Experiment Setup}
\vspace{-1mm}
For training, we use the AED architecture with a Conformer as the encoder, and a shallow single-layer transformer \cite{vaswani2017attention} as the attention-based decoder. 
For \textit{LibriSpeech experiments}, we use a Conformer-12x512x8, which consists of 12 encoder layers with 512 feature dimensions and 8 self-attention heads. We train a 24-layered transformer-based neural LM on the \textit{librispeech-train} dataset to use for rescoring. For the \textit{small-scale experiments} we use a Conformer-16x512x4, whereas for the \textit{large-scale experiments} we use a Conformer-20x512x8. The kernel size of our convolution modules is 31. We optimise our model via the hybrid CTC and attention losses. All of our models are trained using ESPNet \cite{watanabe2018espnet}, with the Adam optimizer \cite{kingma2014adam} and a warm-up learning rate scheduler. 

For front-end, we use 80 dimensional log-mel features and SpecAugment~\cite{park2019specaugment} to perform data augmentation. The BPE embedding is 1024 and 2048 for the small- and large-scale experiments respectively. We train a 4-gram LM on the training text for shallow fusion. For evaluation, we discard the attention-based decoder and only use the CTC decoder to generate outputs with a CTC prefix beam search and beam size of 50. A CTC decoder optimises the real-time factor (RTF) compared to the attention-based decoder, as the latter is non-autoregressive and also needs triggered attention \cite{moritz2019triggered} for streaming inference and is therefore slower. We opt for the CTC decoder as ensuring a low RTF is key for streaming applications.

\vspace{-4mm}
\section{Results}

\vspace{-1mm}
\subsection{Ablation Study on Small-Scale Model}
\label{sec:5k}
\vspace{-1mm}

In the next subsections, we start with discussing the different contributions of our work by analyzing the small-scale results in Table~\ref{table:5k}.

\begin{figure*}[t]
    \centering
    \includegraphics[width=0.86\linewidth]{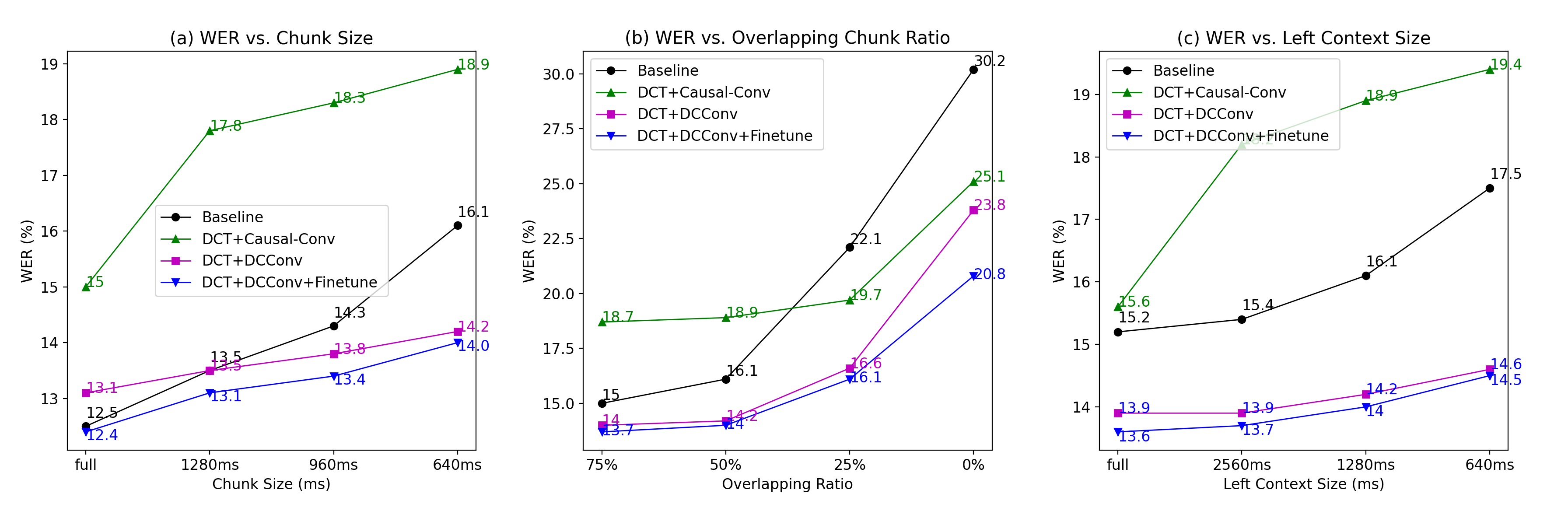}
    \vspace{-2em}
    \caption{\small Ablation study on small-scale models on how the chunk size, overlapping ratio and left context size affect the streaming decoding WER on the Voxpopuli testset. (a) Different chunk sizes with 50\% overlapping and 1280ms left context. (b) Different overlapping ratios with 640ms chunk size and 1280ms left context. (c) Different left context sizes with 640ms chunk size and 50\% overlapping chunk decoding.}
    \label{fig:ablation_hps}
    \vspace{-4mm}
\end{figure*}

\vspace{-3mm}
\subsubsection{DCConv vs Causal and Normal Convolutions}
\vspace{-1mm}

We demonstrate the effectiveness of our novel DCConv by performing an ablation study on three different models: a DCT model with normal convolution (D), a DCT model with causal convolution (E), and a DCT model with our own DCConv (F). Our model outperforms the two baselines on every dataset in the streaming mode. In the non-streaming application on the other hand, our model always beats the DCT model with causal convolution, but does degrade the model with regular convolution. Devising our convolution in such a way that within-chunk future context is used for a more informative acoustic representation (as opposed to (D)), while refraining of using outside-chunk future context to match the training and inference modes more closely (as opposed to (E)), is therefore empirically shown to be advantageous in most cases.

\vspace{-3mm}
\subsubsection{Fine-tuning from Full-Contextual Model}
\vspace{-1mm}

Fine-tuning (G) from a pre-trained full-contextual model instead of training a DCConv model from scratch (F) leads to improvements in both the non-streaming and streaming mode for every single dataset. We observe an average relative WER improvement of 5.5\% and 2.4\% in the non-streaming and streaming modes respectively. Furthermore, with the exception of the WSJ dataset, we now always outperform the regular convolution model in the non-streaming mode. Fine-tuning helps as the model relies on previously gained knowledge from a pre-trained model instead of learning from scratch. It maintains and even improves the non-streaming performance of the full-contextual model, while boosting the streaming performance as it leverages previously learned speech recognition knowledge common to both streaming modes.

\vspace{-3mm}
\subsubsection{P-Conf vs Conformer}
\vspace{-1mm}

Our experiments indicate that using a P-Conf instead of a regular Conformer is beneficial in the streaming mode, while it shows similar results in non-streaming. This holds for the full-contextual (H), DCConv (I) and fine-tuned DCConv (J) models, where using a P-Conf (J) leads to an relative WER improvement of 1.9\%, 0.5\%, 4.2\% and 1.4\% respectively across the 4 datasets over Conformer (G). We hypothesize that the lower receptive field of the P-Conf as a result of the parallel MSA and convolution model during training makes the model more robust in the streaming mode, as it more closely matches the inference conditions.

\vspace{-3mm}
\subsubsection{Ablation Study with Different Streaming Parameters}
\vspace{-1mm}

We take a closer look at the latency vs accuracy trade-off impact with different model parameters that can be controlled during streaming inference. In Fig.~\ref{fig:ablation_hps}, we run an ablation study on the DCT hyper-parameters. Fig.~\ref{fig:ablation_hps}a demonstrates improvement in WER with the increase in chunk size. Likewise, Fig.~\ref{fig:ablation_hps}b and Fig.~\ref{fig:ablation_hps}c reveal better WER performance with more overlap between successive chunks and greater left context size. All are the result of a wider and therefore superior acoustic representation. However, these improvements go at the expense of speed due to the latency-accuracy trade-off. Depending on the use-case we can then select one or the other setting. More importantly, we observe for all settings that our fine-tuned DCConv model performs the best, illustrating the robustness of our approach. In our subsequent experiments, we opt for a 640ms chunk with a 1280ms left context in order to keep those values low to mimic a real-life streaming setting. Furthermore, we stick to a 50\% overlapping ratio as this ratio performs only slightly worse than the 75\% ratio but provides better latency.

\vspace{-3mm}
\subsection{Result with Large-Scale Model}
\label{sec:50k}
\vspace{-1mm}

In Table~\ref{table:5k}, we also compare a fine-tuned DCConv Conformer model (L) to a baseline full-contextual model (K) on a large-scale dataset. The results show no degradation (except for WSJ), and even minor improvements, in the non-streaming mode. This is despite the fact that our model was trained in a unified fashion. In the streaming mode, we observe an average WERR improvement of 14.0\% across all datasets. This validates the gains of our suggested contributions.

\vspace{-3mm}
\subsection{Results on LibriSpeech}
\label{sec:librispeech}
\vspace{-1mm}

\begin{table}[t]
    \centering
    \vspace{-3mm}
    \caption{\small LibriSpeech experiment. Chunk size of 640ms and left context size of 1280ms for 50\% overlapping chunk streaming.
    }
    \vspace{0.6em}
    \resizebox{0.5\textwidth}{!}{
    \begin{tabular}{lcccc}
    \toprule
    Model & \multicolumn{2}{c}{test-clean} & \multicolumn{2}{c}{test-other} \\
    & Non-streaming & Streaming & Non-streaming & Streaming \\\midrule
     (A) Conformer (Full-context) & 2.1 & 3.6 & 5.1 & 9.4 \\
     (B) \hspace{0.5em} + DCT w/ causal conv & 2.6 & 2.9 & 5.8 & 6.8 \\ 
     (C) \hspace{0.5em} + DCT w/ DCConv & 2.3 & 2.6 & 5.4 & \textbf{6.6} \\
     (D) \hspace{1em} + Fine-tune & \textbf{2.0} & \textbf{2.5} & \textbf{4.8} & \textbf{6.6} \\ \midrule
     (E) P-Conformer (Full-context) & 2.1 & 3.4 & 4.9 & 9.1 \\
     (F) \hspace{1em} + DCT w/ DCConv + Fine-tune & \textbf{2.0} & \textbf{2.4} & \textbf{4.8} & \textbf{6.5} \\ \bottomrule
    \end{tabular} }
    \\ [1ex]
    \label{table:librispeech}
    \vspace{-7mm}
\end{table}

In Table~\ref{table:librispeech}, we illustrate the performance of our proposed approach when trained and tested on the widely used public LibriSpeech dataset. First, we compare our DCConv model (C) to a full-contextual model trained without DCT (A) and a DCT model with a regular causal convolution (B) in a non-streaming and a 50\% overlapping streaming mode. We observe a 28.9\% WERR improvement compared to the full-contextual model in the streaming mode, emphasizing the utility of DCT training. Additionally, we notice an average 7.9\% WERR improvement compared to the DCT model with regular causal convolution for both streaming modes, proving the effectiveness of our devised convolution. Furthermore, we demonstrate that fine-tuning (D) instead of training a DCConv model from scratch is especially helpful if you wish to keep a high non-streaming performance of your unified model. It even outperforms the full-contextual model by a WERR improvement of 5.3\% in the non-streaming mode, while keeping the overlapping streaming performance almost intact. Lastly, we observe a minor 2.8\% relative WER improvement when using a P-Conf (F) instead of the conventional one (D) in the streaming mode.

Overall, compared to the full-contextual Conformer model (A), our final fine-tuned DCConv P-Conf model (F) improves the WER by 32.1\% in the streaming mode and reduces the degradation of that mode over the non-streaming full-contextual model from 41.7\% and 45.7\% to 16.7\% and 26.2\% on the \textit{test-clean} and \textit{test-other} datasets respectively. Moreover, we further improve on the state-of-the-art model (B) that reduces this streaming gap too by an average 15.5\% WERR across the 4 settings.

\vspace{-3mm}
\section{Conclusion}
\label{sec:conclusion}
\vspace{-3mm}

In this work, we propose a novel \textit{dynamic chunk convolution} that further improves the existing dynamic chunk training. We achieve this as our convolution better mimics the inference conditions at training time, while keeping a rich acoustic representation. Additionally, we introduce a fine-tuning mechanism and a parallel Conformer block for the unified ASR setting. Our results demonstrate that our unified streaming model closes and even exceeds the gap with a full-contextual model operating in a non-streaming mode, while also showcasing improvements in the streaming mode under different latency constraints. Overall, we outperform the previous state-of-the-art by an average 15.5\% WERR across the LibriSpeech datasets.



\bibliographystyle{IEEEbib}
\bibliography{strings,refs}

\end{document}